\documentclass[aps,prc,showpacs,twocolumn,preprintnumbers,floatfix, tightenlines]{revtex4}
\usepackage{amsmath,amssymb,amsfonts}
\usepackage{hyperref}
\usepackage{ulem}
\usepackage{graphicx}
\usepackage{color}
\usepackage{longtable}

\allowdisplaybreaks

\begin{document}


\title{Constraining the symmetry energy content of nuclear matter \\from nuclear masses: 
a covariance analysis}
\author{C. Mondal}
\address{Saha Institute of Nuclear Physics, 1/AF Bidhannagar, Kolkata
{\sl 700064}, India}
\author{B. K. Agrawal}
\address{Saha Institute of Nuclear Physics, 1/AF Bidhannagar, Kolkata
{\sl 700064}, India}
\author{J. N. De}
\address{Saha Institute of Nuclear Physics, 1/AF Bidhannagar, Kolkata
{\sl 700064}, India}

\begin{abstract}

Elements of nuclear symmetry energy evaluated from different energy density functionals 
parametrized by fitting selective bulk properties of few representative nuclei 
are seen to vary widely. Those obtained from experimental data 
on nuclear masses across the periodic table, however, show that they are better 
constrained. A possible direction in reconciling this paradox may be gleaned from 
comparison of results obtained from use of the binding energies in the fitting 
protocol within a microscopic model with two sets of nuclei, one a representative standard set 
and another where very highly asymmetric nuclei are additionally included.
 A covariance analysis reveals that the additional fitting protocol reduces 
the uncertainties in the nuclear symmetry energy coefficient,
its slope parameter as well as the neutron-skin thickness in $^{208}$Pb
nucleus by $\sim 50\%$. The central values of these entities are also seen
to be slightly reduced.

\end{abstract}
\maketitle


\section{Introduction}
Constraining quantitatively the symmetry energy $C_{v}^0$ of
nuclear matter at saturation density $\rho_0$ and its density slope
$L_0(=3\rho_0\left .\frac{\partial C_v(\rho)}{\partial \rho}\right
.|_{\rho_0})$ has been a major focus of attention in present-day nuclear
physics.  They are fundamental parameters in influencing the binding
energies and stability of atomic nuclei and in shaping the isospin
distributions there giving rise to neutron-skins ($\Delta r_{np}$) in
nuclei of large neutron excess \cite{Myers69,Myers80,Moller12}, where
$\Delta r_{np}$ is the difference between the rms radii for density
distributions of the neutrons and protons in a nucleus. They are also of
seminal importance in astronomical context. The interplay of gravitation
with the pressure of neutron matter $P_n (=3\rho_0 L_0)$ at $\rho_0$ is
a determining factor in the radii of neutron stars \cite{Chen14}. The
dynamical evolution of the core-collapse of a massive star and the
associated explosive nucleosynthesis depend sensitively on the slope
parameter $L_0$ \cite{Steiner05,Janka07}. It also controls the nature
and stability of phases within a neutron star, its critical composition,
thickness and frequencies of crustal vibration \cite{Steiner08} and
determines the feasibility of direct Urca cooling processes within its
interior \cite{Rutel05, Lattimer91, Steiner05}.

The value of the symmetry energy coefficient $C_v^0$ has been known
for some decades to be in the range of $\sim 32.0\pm 4.0$ MeV 
\cite{Myers66, Moller95, Pomorski03, Chang10}. In a controlled 
finite-range droplet model (FRDM) from a fit 
of the observed nuclear masses, the uncertainty in its value shrunk 
appreciably, it was found to be $C_v^0= 32.5 \pm 0.5$
MeV \cite{Moller12}. Studying meticulously the double differences of
 symmetry energies estimated from experimental nuclear masses, Jiang {\it et al} \cite{Jiang12} find
its value to be  $32.10 \pm 0.31$ MeV.  Their procedure takes advantage
of the fact that other effects like pairing and shell corrections
in symmetry energy are well cancelled out from double differences of
neighboring nuclei.

Initial explorations on the symmetry slope $L_0$, however, show wide variations.
The suggested lower limit is $\sim 20$ MeV, on the high side,
it is $\sim 120$ MeV \cite{Centelles09}. 
Astrophysical observations of neutron star radii and
masses provide \cite{Steiner12} a value of $L_0$ between 43 and 52 MeV,
though.  On the other hand, correlation systematics of $\Delta r_{np}$
with nuclear isospin \cite{Centelles09}, isoscaling \cite{Shetty07},
nucleon emission ratios \cite{Famiano06}, isospin diffusion \cite{Li08}
in heavy-ion collision, analysis of giant dipole resonance of $^{208}$Pb
\cite{Trippa08, Roca-Maza13a}, giant quadrupole resonance in $^{208}$Pb 
\cite{Roca-maza13}, pigmy dipole resonance in $^{68}$Ni and $^{132}$Sn
\cite{Carbone10} or of nuclear ground state properties
using the standard Skyrme Hartree-Fock approach \cite{Chen11}, all yield
values of $L_0$ that differ substantially from one another \cite{Tsang12}.
Efforts have also been made to constrain the value of $L_0$ by fitting
the binding energies of large number of deformed nuclei within the
density dependent point coupling model \cite{Niksic08, Zhao10} and by  
considering the pseudo data on the equation of state for the asymmetric 
nuclear matter or pure neutron matter in the fitting protocol of the 
Skyrme models \cite{Chabanat98, Roca-Maza12} as well as the density dependent 
meson exchange models \cite{Roca-Maza11}. These studies indicate that 
$L_0$ is in the range of  40 - 70  MeV.

\begin{table*}[t]
\caption{\label{tab1} The best fit values for the parameters 
of model-I and model-II. 
$m_\sigma$ is 
the mass of $\sigma$ meson given in units of MeV. The masses of $\omega$ and $\rho$ mesons are 
kept fixed to $m_\omega$= 782.5 MeV and $m_\rho$= 763 MeV and nucleon mass is taken to be $M$= 939 MeV.
Uncorrelated errors on the fitted parameters  (upper line) and total 
correlated errors on them (lower line) are given inside the parenthesis for both the models.} 
 \begin{ruledtabular}
\begin{tabular}{ccccccccc}
Name  & $g_{\sigma}$ & $g_{\omega}$ & $g_{\rho}$ & ${\kappa_3}$ & ${\kappa_4}$ &
${\eta_{2\rho}}$ & $\zeta_0$ & $m_{\sigma}$ \\
\hline
model-I & -10.62457 & 13.8585 & 12.077 & 1.46285 &
-0.9673 & 28.33 & 5.2056 & 496.0067 \\
& (0.00013) & (0.0002) & (0.045) & (0.00024) &
(0.0024) & (0.47) & (0.0027) & (0.0062) \\
& (0.246) & (0.662) & (2.60) & (0.275) & (3.66) & (29.9) & (3.21) &
 (12.2)\\
model-II& -10.62123 & 13.85989 & 12.436 & 1.46223 &
-0.8566 & 32.50 & 5.3220 & 495.8146 \\
& (0.00011) & (0.00017) & (0.046) & (0.00021) &
(0.0022) & (0.49) & (0.0027) & (0.0054) \\
& (0.149) & (0.262) & (1.54) & (0.290) & (1.53) & (18.1) & (0.099) &
 (8.23)\\
\end{tabular}
\end{ruledtabular}
\end{table*}

Analysis of data on the nuclear masses in macroscopic models 
seems to contain the fluctuations in the value of $L_0$ somewhat better.
In a macroscopic nuclear model such  as liquid drop, surface tension
tends to favor a nearly uniform drop of neutrons and protons. For a
nucleus, say, with large neutron excess it is energetically advantageous
to distribute some neutrons out at the surface of reduced density
(where the symmetry energy is lower) giving rise to a neutron skin.
Imprints of symmetry energy coefficient $C_v^0$ and its density slope 
$L_0$ are thus encoded in the precisely known nuclear masses.  Fitting the 
few thousand observed nuclear
masses within the FRDM \cite{Moller12}, the value of $L_0$ was found
to be 70$\pm$15 MeV.  
The surface symmetry coefficient $C_s$, along with $C_v^0$ might also be 
used to estimate the value of $L_0$ \cite{Centelles09, Steiner05}. Consistent 
with the values of $C_v^0$ and $C_s$ (as determined from double differences of 
 symmetry energies estimated from experimental nuclear masses \cite{Jiang12} to be 58.91$\pm$1.08 MeV), 
a recent effort aided by microscopic calculations on the $\Delta r_{np}$ of heavy
nuclei \cite{Agrawal12,Agrawal13,Liu13} gives $L_0 = 59\pm 13$ MeV.
This is not far from the one ($L_0 = 65.5 \pm 13.5$ MeV) obtained from
the nuclear equation of state (EoS) constructed \cite{Alam14}
from some well-accepted empirical macroscopic nuclear constraints. 
Exploiting nuclear masses thus seems more or less to constrain better the 
symmetry elements $C_v^0$ and $L_0$ of nuclear matter.
\begin{figure}[h]{}
\resizebox{3.2in}{!}{\includegraphics[]{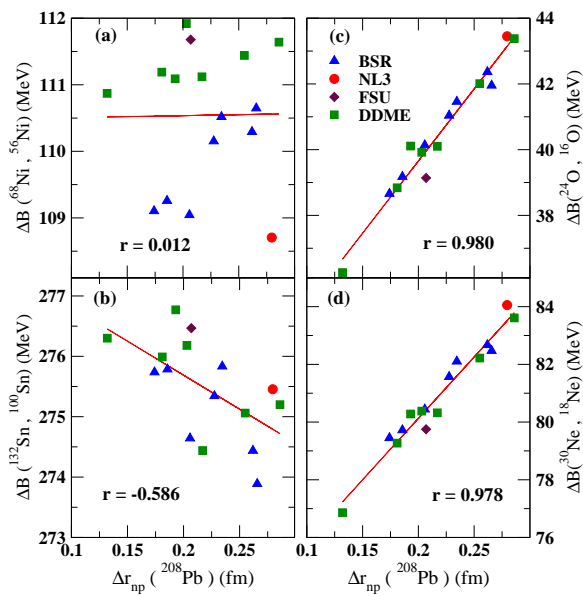}} 
\caption{\label{fig1}
(Color online) The binding energy difference $\Delta B(X,Y)=B(X)-B(Y)$ for 
four different pairs of isotopes are plotted against neutron-skin thickness
$\Delta r_{np}$ in the $^{208}$Pb nucleus for 16 different RMF models (See
text for details). The values of Pearson's correlation coefficients $r$ are 
also displayed. } 
\end{figure}

Energy density functionals (EDF) in microscopic mean field models, parametrized 
to reproduce the binding energies of nuclei alongwith some other specific nuclear 
observables do not, however, display such constraints on the symmetry elements of 
nuclear matter. The EoS constructed out of the different EDFs show wide variations 
in the value of $C_v^0$ and $L_0$ \cite{Dutra12, Dutra14}. Questions then arise how the  
information content of symmetry energy gets blurred in the exploration of nuclear masses 
in microscopic models. For example in the study \cite{Brown00a} of the binding 
energy difference $\Delta B$ of $^{132}$Sn and $^{100}$Sn (which is sensitive to 
the symmetry energy content) vis a vis the neutron skin $\Delta r_{np}$ of 
$^{132}$Sn for different sets of Skyrme EDFs, no noticeable correlation 
between $\Delta B$ and $\Delta r_{np}$ was found. The different EDFs reproduce the 
binding energy difference fairly well, but the value of $\Delta r_{np}$ differ in 
a wide range. 

This quizzical finding leaves open the understanding on the incompatibility of 
the tighter constraints on the 
symmetry energy coefficients of nuclear matter extracted from experimental nuclear 
masses with the larger uncertainties on the values of $C_v^0$ 
and $L_0$ as obtained from parametrizations of the microscopic mean field models. 
The purpose of this communication is to give a general direction in 
shading light on this seeming 
paradox. In an exploratory calculation, we show that the apparent irreconciliation can 
be eased if some highly asymmetric even-even spherical nuclei are additionally included in the fitting 
protocol of the optimization of the EDFs in microscopic mean field models. 
The relativistic  mean-field model
(RMF) is chosen as the vehicle for the realization of our goal.

To keep the matter simple yet pointed, 
we exploit the binding energy difference
between four pairs of nuclei 
in different RMF models
showing increasing 
asymmetry effects: ($^{68}\text{Ni}- ^{56}\text{Ni}$), ($^{132}\text{Sn}-^{100}\text{Sn}$), 
($^{24}\text{O}-^{16}\text{O}$) and ($^{30}\text{Ne}-^{18}\text{Ne}$). 
The neutron rich $^{68}$Ni and $^{132}$Sn nuclei have asymmetries
$\delta$=0.176 and 0.242 respectively ($\delta$ is the isospin asymmetry
parameter $(N-Z)/A$); $^{24}$O and $^{30}$Ne have $\delta$=0.3. 
The Ni and Sn isotopes are doubly closed
shell nuclei. So also the O-nuclei, $^{24}$O is recently seen to be an
unexpectedly stable doubly magic nucleus \cite{Kanungo09, Janssens09}. 
The Ne-nuclei have their neutron shells closed but have valence protons. The
binding energy difference between the two Ne-nuclei is expected to cancel
the pairing and the possible core-polarization effects arising from the
two valence protons partially.
In Fig.\ref{fig1} the binding energy difference between the four pairs of
nuclei are plotted against $\Delta r_{np}$ of $^{208}$Pb, the $\Delta
r_{np}$ and the binding energies being calculated for seven models of
BSR family \cite{Dhiman07, Agrawal10}, NL3 \cite{Lalazissis97}, FSU
\cite{Rutel05} and for seven models of Density Dependent Meson Exchange 
(DDME) family \cite{Vretenar03}.
The correlation coefficient for the Ni-pair is seen to be only 0.012,
for the Sn-pair, it has increased to 0.586. For the O and Ne pairs,
they are quite high, 0.980 and 0.978, respectively. One can not
but fail to notice the increasingly high correlation with increasing
asymmetry, particularly for the latter two cases.

The occurrence of strong correlation for the case of O and Ne pairs 
probably suggest that selective combination of suitable binding energies of
nuclei of low and high isospin may be ideally suited to better constrain
the isovector part of the nuclear interaction.  
A covariance analysis uncovers correlations
among the physical observables; revelation of this interdependence may
be of great importance in assigning   values to the physical observables
of interest with statistical uncertainties. We aim to do that presently, 
the observables  we explore are those that carry the
imprint of isovector content of nuclear interaction i.e.  the symmetry
energy $C_v^0$ and the symmetry energy slope $L_0$ along with the $\Delta
r_{np}$ of $^{208}$Pb.

\section{Effective Lagrangian}
The effective Lagrangian density for the RMF model employed in
the present work is similar to that of the FSU one \cite{Rutel05,
Furnstahl97, Boguta77, Boguta83}. The interaction part of the Lagrangian can be written as,
\begin{eqnarray} \label{eq:Lagrangian2} 
{\cal L}_{\it int}=&&\overline{\psi}\left [g_{\sigma} \sigma -\gamma^{\mu} \left (g_{\omega }
\omega_{\mu}+\frac{1}{2}g_{\mathbf{\rho}}\tau .
\mathbf{\rho}_{\mu}+\frac{e}{2}(1+\tau_3)A_{\mu}\right ) \right ]\psi\nonumber
\\ &&-\frac{{\kappa_3}}{6M}
g_{\sigma}m_{\sigma}^2\sigma^3-\frac{{\kappa_4}}{24M^2}g_{\sigma}^2
m_{\sigma}^2\sigma^4+ \frac{1}{24}\zeta_0
g_{\omega}^{2}(\omega_{\mu}\omega^{\mu})^{2}\nonumber
\\ &&+\frac{\eta_{2\rho}}{4M^2}g_{\omega}^2m_{\rho
}^{2}\omega_{\mu}\omega^{\mu}\rho_{\nu}\rho^{\nu}.  
\end{eqnarray} 
It contains the usual Yukawa coupling between the nucleonic
field ($\psi$) and different meson fields, namely, isoscalar-scalar
$\sigma$ (coupling constant $g_{\sigma}$), isoscalar-vector $\omega$
($g_{\omega}$), isovector-vector $\rho$ ($g_{\rho}$) and photon $A^{\mu}$.
The parameters $\kappa_3$, $\kappa_4$ and $\zeta_0$ determine the strength
of self-couplings of $\sigma$ and $\omega$ mesons. The self-couplings
enable one to produce softening in the EoS of symmetric nuclear
matter if needed.
 The $\omega$-$\rho$ cross-coupling
with coupling-constant $\eta_{2\rho}$ controls the density dependence
of symmetry energy.

\section{Fit data and model parameters}
The values of the parameters entering the EDF of the RMF model are
obtained from an optimal $\chi^2$ fit of the experimental observables
with the theoretically calculated values.  The optimal values of
the parameters of the EDF give the minimum $\chi_0^2({\bf p})$ of
the objective function $\chi^2({\bf p})$ (${\bf p}$ referring to
the parameter space), defined as, 
\begin{equation} 
\chi^2({\bf p}) =\frac{1}{N_d - N_p}\sum_{i=1}^{N_d} \left (\frac{ \mathcal{O}_i^{exp} - \mathcal{O}_i^{th}({\bf p})}{\Delta\mathcal{O}_i}\right )^2,
\label {chi2} 
\end{equation} 
where, $N_d$ and $N_p$ are the number
of  experimental data points and the number of fitted  parameters,
respectively. $\Delta\mathcal{O}_i$ is the adopted error and
$\mathcal{O}_i^{exp}$ and $\mathcal{O}_i^{th}({\bf p})$ are the
experimental and the corresponding theoretical values for a given
observable. 
The values of  $\Delta\mathcal{O}_i$ are chosen in such a way that 
$\chi_0^2({\bf p})$ is unity.
The experimental observables, in our case, include only
the values of the binding energies and charge radii of a selected set
of nuclei.  There may be values of parameters that are also likely, they
can give a reasonable fit to the data, the reasonableness of the domain
of parameter space being defined as $\chi^2({\bf p}) \le \chi^2_0({\bf
p}) + 1$ \cite{Brandt97, Reinhard10, Dobaczewski14}. A reasonable
domain of parameters yields a reasonable set of results for each set
of predicted observables. 

\begin{table*}[t]
\caption{\label{tab2}
Observables $\mathcal{O}$ of different nuclei, adopted errors on them 
$\Delta\mathcal{O}$, their experimental values and the ones obtained for 
 model-I and II. $B$ and $r_{ch}$ refers to binding energy and charge radius of 
a nucleus respectively, and $\Delta B$ is binding energy difference of two isotopes of 
a nucleus as indicated. $B$ and $\Delta B$ are in units of MeV and $r_{ch}$ in fm.
} 
 \begin{ruledtabular}
\begin{tabular}{cccccc}
Nucleus & $\mathcal{O}$ &$\Delta\mathcal{O}$ & Expt. &  
model-I & model-II \\
\hline
$^{16}$O & $B$ &4.0 & 127.62 &  127.781$\pm$0.990 & 127.783$\pm$0.576 \\
& $r_{ch}$ &0.04 & 2.701 &  2.700$\pm$0.017 & 2.699$\pm$0.013 \\
$^{16}$O, $^{24}$O & $\Delta B$ &2.0 & 41.34 &  - & 40.995$\pm$1.046 \\
$^{18}$Ne, $^{30}$Ne & $\Delta B$ &2.0 & 79.147 &  - & 79.149$\pm$1.296 \\
$^{40}$Ca & $B$ &3.0 & 342.051 &  342.929$\pm$1.064 & 342.927$\pm$0.927 \\
& $r_{ch}$ &0.02 & 3.478 &  3.457$\pm$0.013 & 3.455$\pm$0.010 \\
$^{48}$Ca & $B$ &1.0 & 415.99 &  414.883$\pm$0.720 & 414.751$\pm$0.541 \\
& $r_{ch}$ &0.04 & 3.479 &  3.439$\pm$0.007 & 3.439$\pm$0.006 \\
$^{56}$Ni & $B$ &5.0 & 483.99 &  483.752$\pm$2.495 & 483.619$\pm$1.646 \\
& $r_{ch}$ &0.18 & 3.750 &  3.695$\pm$0.025 & 3.693$\pm$0.020 \\
$^{68}$Ni & $B$ &1.0 & 590.43 &  592.294$\pm$0.784 & 592.162$\pm$0.736 \\ 
$^{90}$Zr & $B$ &1.0 & 783.893 &  782.855$\pm$4.833 & 782.776$\pm$1.621 \\
& $r_{ch}$ &0.02 & 4.269 &  4.267$\pm$0.009 & 4.267$\pm$0.034 \\
$^{100}$Sn & $B$ &2.0 & 825.8 &  827.987$\pm$1.753 & 827.757$\pm$1.534 \\
$^{116}$Sn & $B$ &2.0 & 988.32 &  987.169$\pm$0.946 & 987.072$\pm$0.760 \\
& $r_{ch}$ &0.18 & 4.626 &  4.623$\pm$0.009 & 4.623$\pm$0.008 \\
$^{132}$Sn & $B$ &1.0 & 1102.9 &  1102.851$\pm$1.146 & 1102.631$\pm$0.856 \\
& $r_{ch}$ &0.02 & 4.71 &  4.711$\pm$0.011 & 4.712$\pm$0.010 \\
$^{144}$Sm & $B$ &2.0 & 1195.74 &  1195.834$\pm$1.240 & 1195.736$\pm$1.287 \\
& $r_{ch}$ &0.02 & 4.96 &  4.956$\pm$0.009 & 4.956$\pm$0.009 \\
$^{208}$Pb & $B$ &1.0 & 1636.446 &  1636.457$\pm$4.301 & 1636.383$\pm$0.917 \\
& $r_{ch}$ &0.02 & 5.504 &  5.530$\pm$0.012 & 5.531$\pm$0.010 \\
\end{tabular}
\end{ruledtabular}
\end{table*}

To explore 
the results of Fig.\ref{fig1} more critically, 
two models (model-I and model-II) corresponding to
different sets of fit-data are constructed.
 In model-I the binding energies and charge radii of some standard set
 of nuclei
($^{16}$O, $^{40}$Ca, $^{48}$Ca, $^{56}$Ni, $^{68}$Ni, $^{90}$Zr,
$^{100}$Sn, $^{116}$Sn, $^{132}$Sn, $^{144}$Sm and $^{208}$Pb)
spanning the entire periodic table are taken as fit-data.  
In model-II, we have the same set
of experimental observables, but with the addition of the binding energy
difference $\Delta B$ of ($^{24}\text{O},^{16}\text{O}$) and of ($^{30}\text{Ne},^{18}\text{Ne}$).
The parameters of model-I and model-II are obtained by optimizing \cite{Bevington69} 
the objective function $\chi^2$({\bf p}) as given by Eq. (\ref{chi2}).

Once the optimized parameter set is obtained the correlation coefficient 
between two quantities $\mathcal{A}$ and $\mathcal{B}$, which may be a parameter as well as an 
observable, can be evaluated within the covariance analysis as,
\begin{equation}
  {c}_\mathcal{AB} =
  \frac{\overline{\Delta \mathcal{A}\,\Delta \mathcal{B}}}
       {\sqrt{\overline{\Delta \mathcal{A}^2}\;\overline{\Delta \mathcal{B}^2}}},
\label{corr}
\end{equation}
where, covariance between $\mathcal{A}$ and $\mathcal{B}$ is expressed as, 
\begin{equation}
\overline{\Delta \mathcal{A}\,\Delta \mathcal{B}}=\sum_{\alpha\beta}\left(\frac{\partial \mathcal{A}}{\partial
\rm{p}_{\alpha}}\right)_{\bf p_0}\mathcal{C}_{\alpha\beta}^{-1}\left(\frac{\partial \mathcal{B}}{\partial \rm{p}_{\beta}}
\right)_{\bf p_0}. 
\label{deltaab}
\end{equation}
Here, $\mathcal{C}_{\alpha\beta}^{-1}$ is an element of inverted curvature matrix given by,
\begin{equation}
 \mathcal{C}_{\alpha\beta}=\frac{1}{2}\Big(\frac{\partial^2 \chi^2(\mathbf{p})}{\partial {\rm {p}_{\alpha}}\partial {\rm{p}_{\beta}}}\Big)
_{\mathbf{p}_0},
\label{Cmatrix}
\end{equation}
where, {\bf p}$_0$ represents the optimized set of parameters. The square
of the error, ${\overline{\Delta \mathcal{A}^2}}$ in $\mathcal{A}$ can be computed
using Eq. (\ref{deltaab}) by substituting $\mathcal{B=A}$.  Curvature matrix
can also facilitate to locate the reasonable domain of parameters
\cite{Chen14, Dobaczewski14, Erler14}.  
The parameters  for model-I and II corresponding
to minimum value of the objective function $\chi^2$({\bf p})
(=$\chi_0^2$({\bf p})) along with their uncorrelated and correlated statistical 
errors are listed in Table \ref{tab1}.  
The correlated and  uncorrelated errors are computed with and without 
contributions from the non-diagonal elements  of the matrix $\mathcal{C}$ as given
by Eq. \ref{Cmatrix}, respectively.
The correlated errors on the parameters  for the case of model-II are
smaller than those obtained for the model-I indicating that the inclusion
of the fit data on the binding energy differences constrain the model
parameters better. In particular, the errors on the parameters $g_\rho$
and $\eta_{2\rho}$, which govern the isovector part of the effective
Lagrangian, are smaller for the model-II. The large error on the parameters
$\kappa_3$ and $\kappa_4$ for both the models may be due to the fact that the fit data
does not include any observable which could constrain the value of the
nuclear matter incompressibility coefficient \cite{Boguta83}.
In Table \ref{tab2} different observables $\mathcal{O}_i$, adopted errors on them 
$\Delta\mathcal{O}_i$, their experimental values along with the 
results obtained for model-I and model-II using the corresponding best
fit parameters are listed. The values of $\mathcal{O}_i$ and $\Delta \mathcal{O}_i$, except
for the $\Delta B$ of ($^{24}\text{O},^{16}\text{O}$) and ($^{30}\text{Ne},^{18}\text{Ne}$)
and $r_{ch}$ of $^{132}$Sn, are exactly same as used in Ref. \cite{Klupfel09}. The
experimental data for $\Delta B$ of ($^{24}\text{O},^{16}\text{O}$) and
($^{30}\text{Ne},^{18}\text{Ne}$) are taken from \cite{Audi12} and that for the $r_{ch}$ of 
$^{132}$Sn from \cite{Angeli13}.

\section{Results and Discussions}
\begin{figure}[t]{}
\resizebox{3in}{!}{\includegraphics[]{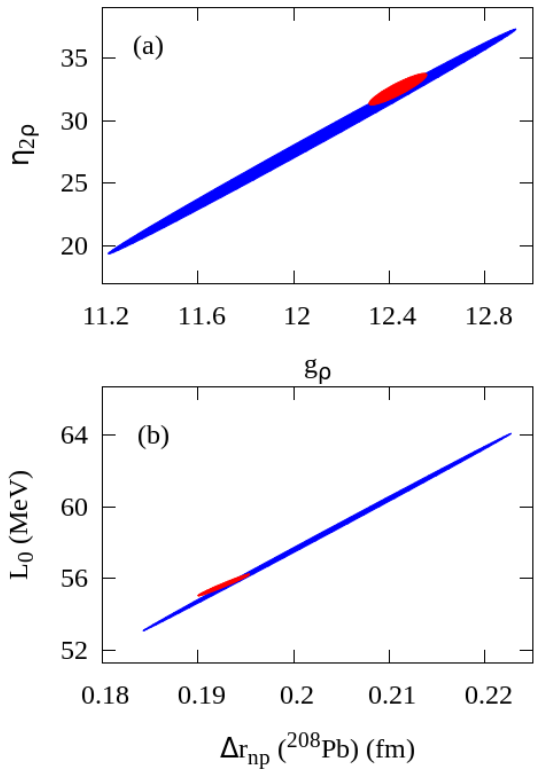}}
\caption{\label{fig2} (Color online) 
The covariance ellipsoids for the parameters $g_\rho \text{ - } \eta_{2\rho}$
(upper panel) and the corresponding $L_0 \text{ - } \Delta r_{np}$ (lower panel) for
the model I (color blue) and model II (color red).
The area inside the ellipsoids indicate the reasonable domain of the
parameters.}
\end{figure}

We now compare the results obtained for the model-I and model-II to see
upto what extent the inclusion of the experimental data on the binding
energy differences between the pair of  O and Ne nuclei can constrain
the iso-vector part of the effective Lagrangian.
In 
Fig. \ref{fig2}(a) the reasonable domain for the
parameters $g_\rho$ and $\eta_{2\rho}$ is displayed.
\begin{table} 
\caption{\label{tab3}
The values for the binding energy
per nucleon $E/A$, incompressibility coefficient $K$, 
Dirac effective mass of nucleon $m^*/m$, symmetry energy coefficient $C_v^0$ and density
slope parameter of symmetry energy $L_0$ for the nuclear matter evaluated 
at saturation density $\rho_0$ along with the correlated errors on them obtained within 
the covariance analysis for the models I and II. The results for
neutron-skin thickness $\Delta r_{np}$ in  $^{48}$Ca, $^{132}$Sn and $^{208}$Pb are also presented.
}
\begin{ruledtabular}
\begin{tabular}{ccc} Observable  & model-I & model-II\\ \hline $E/A$
(MeV)           &  $-16.036\pm0.070$  & $-16.036\pm0.051$   \\ $K$
(MeV)             &  $210.12\pm27.87$   & $209.64\pm28.52$    \\
$\rho_0$ (fm$^{-3}$)  &  $0.150\pm0.003$    & $0.150\pm0.003$     \\
$m^*/m$               &  $0.585\pm0.012$    & $0.585\pm0.010$     \\
$C_v^0$ (MeV)         &  $32.03\pm3.08$     & $31.69\pm1.51$      \\
$L_0$ (MeV)           &  $57.62\pm17.08$    & $55.63\pm7.00$      \\
$\Delta r_{np}$ ($^{48}$Ca) (fm) & $0.191\pm0.036$ & $0.187\pm0.016$\\ 
$\Delta r_{np}$ ($^{132}$Sn) (fm) & $0.266\pm0.070$ & $0.257\pm0.031$\\ 
$\Delta r_{np}$ ($^{208}$Pb) (fm) & $0.201\pm0.065$ & $0.193\pm0.030$\\ 
\end{tabular} 
\end{ruledtabular} 
\end{table}
For this reasonable domain of the parameters the
values of the symmetry energy slope parameter $L_0$ and the $\Delta
r_{ np}$  in the $^{208}$Pb nucleus are displayed in 
Fig. \ref{fig2}(b).
The inclined and elongated shapes of the ellipsoids indicate that the
 correlations amongst $g_\rho \text{ - } \eta_{2\rho}$ and $L_0 \text{ - } \Delta r_{np}$ 
are strong. In fact, the values of the correlation coefficients (Eq.
(\ref{corr})) for these pairs of quantities for both the models turn
out to be $\sim$0.95.  It is evident that the ellipsoids depicting the
results for the model-II (color red) are narrower in comparison  to those for the
model-I (color blue). This is suggestive of the fact that the inclusion of the binding
energies for the $^{24}$O and $^{30}$Ne put tighter constraints on the
isovector part of the effective Lagrangian density.

Nuclear matter properties for model-I and model-II are compared in
Table \ref{tab3}. 
 Errors on the entities describing the isoscalar
behavior of nuclear matter ($E/A$, $K$, $\rho_0$ and $m^*/m$) are pretty
much the same for both the models concerned.  For model-II, however, a
significant improvement (by a factor $\sim2$) on the spread of parameters
like $C_v^0$ and $L_0$, which describe the symmetry behavior of nuclear
matter, is achieved over model-I.
Strikingly, the errors on $C_v^0$ and $L_0$ for the model-II agree very
well with the ones obtained for the SAMi Skyrme force \cite{Roca-Maza12}
which includes the variational EoS for the pure neutron matter as pseudo
data  in the fitting protocol.
We also provide the values of $\Delta r_{np}$ for $^{48}$Ca, $^{132}$Sn
and $^{208}$Pb nuclei in Table \ref{tab3}.  The reduction in the errors on
$\Delta r_{np}$ for the model-II in comparison to those for the model-I
are in harmony with the results depicted in Fig.  \ref{fig2}.

To this end, it may be mentioned that the correlations of neutron-skin
thickness in $^{208}$Pb nucleus with $L_0$ has been studied within the
covariance approach for different nuclear energy density functionals
\cite{Reinhard13}. These correlations exhibit some degree of model
dependence.  In view of this, the present investigation should be extended
to other energy density functionals.
The value of $L_0$ might also be constrained by including in the
fitting protocol the experimental data for the iso-vector giant
dipole polarizability for the $^{208}$Pb nucleus \cite{Reinhard10}. However, it must
be reminded that the  $L_0$ is  well correlated with the product of
dipole polarizability and the symmetry energy coefficient at
the saturation density $C_v^0$ rather than the dipole polarizability alone
\cite{Roca-Maza13a}.

\section{Summary and Conclusion}
To summarize, we have tried to give a critical look on the question why the 
information content on the elements of symmetry energy do not get transported 
properly in the EDFs when they are parametrized to reproduce the binding energies 
of representative nuclei in a mean-field model. The asymmetry parameters are then 
less constrained. Working in the confines of RMF model, we find that better constraint 
can be achieved in the isovector part of the effective Lagrangian density and hence 
in the said symmetry elements if one includes, in the fitting protocol of the 
microscopic model binding energies of some highly asymmetric nuclei. In this 
preliminary investigation, comparison of results with two sets of nuclei, one 
a standard set and another with two additional nuclei of very high asymmetry ($\delta 
= 0.3$) shows a pointer in this direction.
The covariance analysis yields
the reduction in the uncertainties on the values of symmetry energy and its
slope parameter by $\sim50\%$ in comparison to those obtained without the
inclusion of the additional nuclei in the fit data. Similar effects are also observed 
on the values of $\Delta r_{np}$ in $^{48}$Ca, $^{132}$Sn and 
$^{208}$Pb nuclei. A more detailed analysis along this direction 
is underway.



\end{document}